\documentclass[fleqn,twoside]{article}
\usepackage{espcrc2}
\usepackage{epsfig}

\usepackage{graphicx}
\usepackage[figuresright]{rotating}

\newcommand{\AmS}{{\protect\the\textfont2
  A\kern-.1667em\lower.5ex\hbox{M}\kern-.125emS}}

\title{\vspace{-0.8cm}
Partons and QCD in high-energy polarized pp collisions\\
\vspace{-1.9cm}
\hspace*{12.62cm}
\small{BNL-NT-01/23 \\
\hspace*{12.8cm}RBRC-196}
\normalsize
\vspace{1.2cm}
}

\author{M.\ Stratmann\address{Institut f\"ur Theoretische Physik, 
Universit\"at Regensburg,\\
D-93040 Regensburg, Germany}, 
W. Vogelsang\address{Physics Department and RIKEN-BNL Research Center,
Brookhaven National 
Laboratory,\\ Upton, New York 11973, U.S.A.}
\thanks{We are grateful to RIKEN, Brookhaven National Laboratory and the U.S.
Department of Energy (contract number DE-AC02-98CH10886) for
providing the facilities essential for the completion of this work.}
\thanks{Invited talk presented by W. Vogelsang at the
``International Workshop on the Spin Structure of the Proton and
Polarized Collider Physics'', ECT$^*$, Trento, Italy, 
July 23-28, 2001.}}
       
\begin{document}
\renewcommand{\textfraction}{0}
\def    \beq             {\begin{equation}}
\def    \eeq             {\end{equation}}
\def    \beqa            {\begin{eqnarray}}
\def    \eeqa            {\end{eqnarray}}
\def \e {{\rm e}}
\def \as {{\alpha_s}}

\begin{abstract}
We present and use a technique for implementing in a fast 
way, and without any approximations, higher-order calculations 
of partonic cross sections into global analyses of parton distribution 
functions. The approach, which is set up in Mellin-moment space, is 
particularly suited for analyses of future data from polarized 
proton-proton collisions at RHIC, but not limited to this case.
We also briefly discuss the effects of soft-gluon resummations
on spin asymmetries.
\end{abstract}

\maketitle
%
\section{Introduction}
%
\noindent
Current and future dedicated high-energy spin experiments are expected to 
vastly broaden our understanding of the nucleon spin structure.
Processes in polarized lepton-nucleon and proton-proton scattering will be 
studied~\cite{ep,rhic} that will give access to, in particular, the 
spin-dependent gluon distribution $\Delta g$ and the flavor-SU(2) breaking 
in the polarized nucleon sea, $\Delta \bar{u}-\Delta \bar{d}$. 
Having available at some point in the near future spin data on 
various different reactions, one needs to tackle the question 
of how to determine such polarized parton densities from the measurements.
This is not at all a new problem: in the unpolarized
case, several groups perform such ``global analyses'' of the
plethora of data available there~\cite{cteq,grv}. 
The strategy is in principle clear: an ansatz for the parton distributions 
at some initial scale $\mu_0$, given in terms of appropriate functional 
forms with a set of free parameters, is evolved to a scale 
$\mu_F$ relevant for a certain data point for a certain cross section. 
Then the parton densities at scale $\mu_F$ are used to compute the 
theoretical prediction for the cross section, 
and a $\chi^2$ value is assigned that represents the quality of the 
comparison to the experimental point. This is done for all data 
points to be included in the analysis, and subsequently the parameters 
in the ansatz for the parton distribution functions are varied, until 
eventually a minimum in $\chi^2$ is reached. 

In practice, this approach is difficult if the partonic scattering 
is treated beyond the lowest order of perturbation theory (PT).  
The numerical evaluation of the hadronic cross section at higher 
orders is usually a rather time-consuming procedure as it often requires 
several tedious numerical integrations, not only for the 
convolutions with the parton densities, 
but also for the phase space integrations in the partonic
cross section. The fitting procedure outlined above, on the other
hand, usually requires thousands of computations of the cross 
section for any given data point, and so the computing 
time required for a fit easily becomes excessive even on modern
workstations. 

In the unpolarized case, a way to get around this problem is 
based on the fact that the parton densities are already known 
here rather accurately~\cite{cteq,grv}. 
As a consequence, the theory answer for a certain
cross section is expected to change in a very predictable way 
when going from the lowest-order Born level to the 
first-order approximation.
It is then possible to pre-calculate a set of correction factors $K_i$
($i$ running over the data points), 
and to simply multiply them in each step of the fitting procedure to
the {\em lowest-order} approximation for the cross section, the latter 
being usually much faster to evaluate than that involving 
higher order terms. In the polarized case, however, it is in general 
not at all clear whether such a strategy will work. Here, the parton 
densities are known with {\em much} less accuracy so far.  
It is therefore not possible to pre-calculate higher-order correction 
factors that one would be able to keep fixed throughout the fit, while using
``fast'' lowest-order expressions for the partonic cross sections. 
For instance, the spin-dependent parton distributions, 
as well as the polarized partonic cross sections, may have zeros in the 
kinematical regions of interest, near which the predictions at
lowest order and the next order will show marked differences.

Our goal is to find a way
of implementing efficiently, and without approximations, 
higher-order expressions for any hadronic cross section 
into the fitting procedure. 
As was recently shown~\cite{sv}, this can be achieved
in a very simple and straightforward way by going to Mellin-$n$
moment space. A related technique was first discussed in
\cite{bghv,ref:kosower}. 

In the following we will outline our method and apply it to an
example, the production of prompt photons at 
high transverse momentum in polarized $pp$ collisions at RHIC.
The sensitivity of this reaction to the gluon distribution via the 
LO Compton subprocess is the main reason why this process will be the flagship
measurement of $\Delta g$ at RHIC~\cite{rhic}.
As a first case study for future global analyses 
we also carry out a ``toy'' analysis of polarized 
deep-inelastic scattering (DIS)~\cite{emc} {\em and}
prompt photon data projected for RHIC.

At the end of this paper, we digress from our main 
topic and briefly discuss the 
effects of soft-gluon resummation on spin asymmetries.

\section{Hadronic Cross Sections and the Mellin Moment Technique}
\noindent
A general spin-dependent cross section in longitudinally polarized 
$pp$ collisions, differential in a certain observable $O$ and integrated 
over experimental bins in other kinematical variables $T$, can be 
written as
\begin{eqnarray} 
\label{eq1}
\frac{d\Delta \sigma^H}{dO} &\equiv&
\frac{1}{2} \left[\frac{d\sigma^H}{dO}(++) - 
\frac{d\sigma^H}{dO}(+-)
\right]  \nonumber\\[2mm]
&&\hspace*{-1.5cm}= 
\sum_{a,b}\, \int_{\mathrm bin} dT\, 
\int_{x_a^{\mathrm min}}^1 dx_a 
\int_{x_b^{\mathrm min}}^1 dx_b 
\nonumber \\
&&\hspace*{-1.2cm} \times \;\Delta f_a (x_a,\mu_F) \,\Delta f_b (x_b,\mu_F) 
\, \nonumber \\ 
%
&&\hspace*{-1.2cm}
\times \,\frac{d\Delta \hat{\sigma}_{ab}^{H}}{dO dT}
(x_aP_A, x_bP_B, P_H, T, \mu_R, \mu_F) \; ,
\end{eqnarray}
where the arguments $(++)$ and $(+-)$ in the first line
refer to the helicities of the incoming protons. The
$\Delta f_i$ are the spin-dependent parton distributions,
defined as
\begin{equation}\label{eq2}
\Delta f_i(x,\mu_F) \equiv f_i^+(x,\mu_F) -  f_i^-(x,\mu_F) \; ,
\end{equation}
where $f_i^+$ ($f_i^-$) denotes the number density of a parton-type 
$f_i$ with helicity `+' (`$-$') in a proton with positive helicity,
carrying the fraction $x$ of the proton's momentum. Note that for 
some observables, a fragmentation function may be additionally 
present in Eq.~(\ref{eq1}).

The scale $\mu_F$ is the factorization scale for
initial state collinear singularities and reflects the certain 
amount of arbitrariness in the separation 
of short-distance and long-distance physics embodied in Eq.~(\ref{eq1}).
The other scale, $\mu_R$, in Eq.~(\ref{eq1})
is the renormalization scale, introduced in the procedure of 
renormalizing the strong coupling constant. Finally, 
the sum in Eq.~(\ref{eq1}) is over all 
contributing partonic channels $a+b\to
H + X$, with 
$d\Delta \hat{\sigma}_{ab}^{H}$ the associated spin-dependent partonic cross
section, defined in complete analogy with the first line of 
Eq.~(\ref{eq1}). They are perturbative and hence are expanded 
in powers of $\alpha_s$.

The Mellin moments of the polarized parton distribution functions
are defined as
\beq
\Delta f_i^n(\mu) \equiv \int_0^1 dx \;x^{n-1}  \Delta f_i(x,\mu) \; .
\eeq
The parton distributions in Bjorken-$x$ space are recovered from the
moments by an inverse Mellin transform: 
\beq \label{eq6}
 \Delta f_i(x,\mu) = \frac{1}{2 \pi i} \int_{{\cal C}_n} dn \; 
x^{-n} \Delta f_i^n(\mu) \; , 
\eeq
where ${\cal C}_n$ denotes a contour in the complex $n$ plane that 
has an imaginary part ranging from $-\infty$ to $\infty$ and that 
intersects the real axis to the right of the rightmost poles of 
the $\Delta f_i^n(\mu)$. 

The crucial, but simple, step in applying moment techniques to
Eq.~(\ref{eq1}) is to express the $\Delta f_i(x_i,\mu_F)$
by their Mellin inverses in Eq.~(\ref{eq6})
\cite{ref:kosower,sv}. One subsequently interchanges integrations 
and arrives at
\begin{eqnarray} \label{eq7}
\frac{d\Delta \sigma^H}{dO}
&=& 
%
\sum_{a,b} \, \int_{{\cal C}_n} \frac{dn}{2 \pi i} \; 
\int_{{\cal C}_m} \frac{dm}{2 \pi i}
\;\Delta f_a^n\,\Delta f_b^m
\nonumber \\
&& 
\hspace*{-1.5cm}\times \, \int_{\mathrm{bin}} dT\, 
\int_{x_a^{\mathrm min}}^1 dx_a 
\int_{x_b^{\mathrm min}}^1 dx_b \,\,
x_a^{-n} x_b^{-m}
\nonumber \\
&& 
\hspace*{-1.5cm}\times \,
\frac{d\Delta \hat{\sigma}_{ab}^{H}}{dOdT}
(x_aP_A,\, x_bP_B\, ,P_H,T,\,\mu_R,\mu_F) \nonumber \\
&&\hspace*{-1.5cm}\equiv\,
\sum_{a,b} \, \int_{{\cal C}_n} dn \; 
\int_{{\cal C}_m} dm\;\Delta f_a^n(\mu_F)\,\Delta f_b^m(\mu_F)
\nonumber \\
&& 
\hspace*{-1cm}\times \,
\Delta \tilde{\sigma}_{ab}^{H} (n,m,O,\mu_R,\mu_F)
\; .\label{eq8}
\end{eqnarray}
One can now pre-calculate the quantities 
$\Delta \tilde{\sigma}_{ab}^{H} (n,m,O,\mu_R,\mu_F)$, 
which do not depend at all on the
parton distribution functions, {\em prior} to the fit for a specific
set of the two Mellin variables $n$ and $m$, for each contributing 
subprocess and in each experimental bin. Effectively, one has to 
compute the cross sections with complex ``dummy'' parton distribution 
functions $x_a^{-n} \, x_b^{-m}$. All the tedious 
and time-consuming integrations are already dealt with in the calculation 
of the $\Delta \tilde{\sigma}_{ab}^{H} (n,m,O,\mu_R,\mu_F)$.

\begin{figure}[h]
\vspace*{-1cm} 
\hspace*{-1.2cm}
\epsfig{file=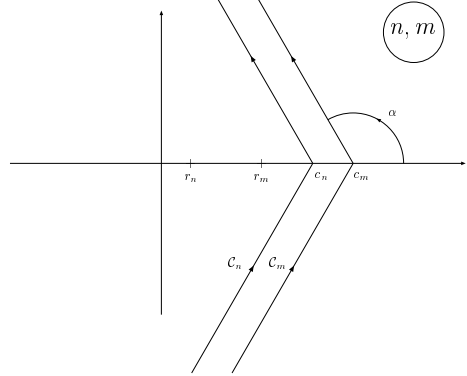,width=8.0cm}
\vspace*{-6.0cm}
\caption{\sf Contours in complex Mellin-$n,m$ spaces for the 
calculation of the double Mellin inverse in Eq.~(\ref{eq8}). 
$r_n$ and $r_m$ denote the rightmost poles of the integrand in $n$ and $m$,
respectively, and the $c_i$ the intersections with the real axis. 
\label{fig1}}
\end{figure}

The double inverse Mellin transformation which finally links the 
parton distributions with the pre-calculated 
$\Delta \tilde{\sigma}_{ab}^{H} (n,m,O,\mu_R,\mu_F)$
of course still needs to be performed 
{\em in each step} of the fitting procedure. However,
the integrations over $n$ and $m$ in Eq.~(\ref{eq8}) 
are extremely fast to perform by  
choosing the values for $n,m$ in  
$\Delta \tilde{\sigma}_{ab}^{H} (n,m,O,\mu_R,\mu_F)$
on the contours ${\cal C}_n$,
${\cal C}_m$ simply as the supports for a Gaussian integration. 
The point here is that the integrand in $n$ and $m$ falls off very 
rapidly as $|n|$ and $|m|$ increase along the contour, for two reasons:
first, each parton distribution function is expected to fall off at
least as a power $(1-x)^3$ at large $x$, which in moment space
converts into a fall-off of $\sim 1/n^4$ or higher. Secondly, 
we may choose contours in moment space that are bent by an 
angle $\alpha-\pi/2$ with respect to the vertical direction;
a possible choice is shown in Fig.~\ref{fig1}. Then, for large
$|n|$ and $|m|$, $n$ and $m$ will acquire large negative real parts, 
so that $(x_a)^{-n}$ and $(x_b)^{-m}$ decrease
exponentially along the respective contours. 
This helps for the numerical convergence of the calculation 
of the $\Delta \tilde{\sigma}_{ab}^{H} (n,m,O,\mu_R,\mu_F)$
and also gives them a rapid fall-off at large 
arguments. 

\vspace*{-0.6cm}
\section{Example: Prompt Photons at RHIC}
%
In Ref.~\cite{sv} we have presented two practical applications of the Mellin 
technique: the semi-inclusive production of hadrons in 
polarized DIS, and the production of prompt photons in 
polarized $pp$ collisions at RHIC. Here we will briefly recall
our findings for the latter process.
To be specific, we consider the transverse momentum ($p_T$) distribution
of the prompt photon, integrated over a certain experimental bin in its
pseudorapidity $\eta$. Thus, we have ``$O\equiv p_T$'' and ``$T\equiv \eta$'' 
in Eq.~(\ref{eq1}).

For prompt photon production, the lowest-order partonic reactions 
are $q+\bar{q}\to \gamma +g$ and $q+g\to \gamma +q$, the latter channel 
being sensitive to the polarized gluon distribution.
The next-to-leading order (NLO) 
corrections are also available~\cite{dgnlo} and will be 
used in our case study.

We use $\sqrt{S}=200$ GeV and consider five 
values of $p_T$ which will be experimentally accessible at RHIC, 
$p_T=[12.5, 17.5, 22.5, 27.5, 32.5]$ GeV. We average 
over $|\eta|<0.35$ in pseudorapidity and impose an isolation cut on 
the photon~\cite{rhic}, for which we choose the criterion
proposed in~\cite{frixione} with parameters $R=0.4$, $\epsilon=1$. 
A positive feature of this isolation criterion is the 
absence of a fragmentation contribution to prompt photon production.
We choose the renormalization and factorization scales $\mu_R=\mu_F= p_T$. 

\begin{figure}[h] 
\hspace*{0cm}
\epsfig{file=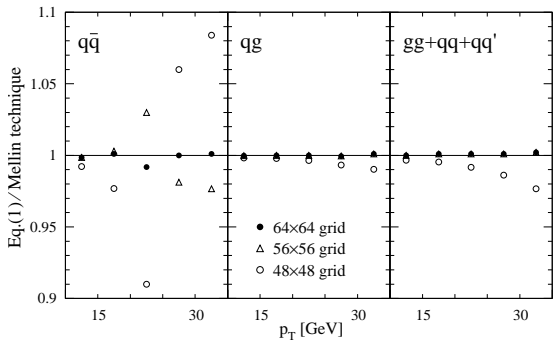,width=8.0cm}
\vspace*{-1.2cm}
\caption{\sf Comparison of the results based on the Mellin-technique in
Eq.~(\ref{eq1}) to those of Eq.~(\ref{eq8}) 
for various sizes of the grid in $n,m$.
\label{fig2}}
\vspace*{-0.5cm}
\end{figure}

Our first goal here is to show that the method based on Eq.~(\ref{eq7}) 
actually works and correctly reproduces the 
result obtained within the direct, but ``slow'', calculation via 
Eq.~(\ref{eq1}). Also, we need to establish an optimal size of the grids 
that yields excellent accuracy but is still calculable in, say, a few 
hours of CPU time on a standard workstation. Fig.~\ref{fig2} 
compares the results based on Eq.~(\ref{eq7}),
referred to as ``Mellin technique'', to those of Eq.~(\ref{eq1}), 
for various sizes of the grid in $n,m$. Here we have used the 
polarized parton densities of~\cite{grsv} (``standard'' set).
\vspace*{-1.cm}
For a more detailed comparison, we split up
the contributions to the NLO prompt photon cross section into three
parts, associated with the reactions $q+\bar{q}\to \gamma +X$ and $q+g\to
\gamma +X$ that are already present at Born level, and all other
processes that arise only at NLO. One notices that in each case already 
a grid size of $64\times 64$ values yields excellent accuracy. Even a
$56\times 56$ grid is acceptable apart from a minor deviation 
occurring for $q\bar{q}$ scattering in the vicinity of a zero in the
partonic cross section.

The crucial asset of the Mellin method is the speed at which one can calculate 
the full hadronic cross section, once the grids 
$\Delta \tilde{\sigma}_{ab}^{\gamma} (n,m,p_T,\mu_R,\mu_F)$
have been pre-calculated. For the $64 \times 64$ grid, we found that
1000 evaluations of the full NLO prompt photon cross section
take only about 10-15 seconds on a standard workstation. Note that 
this number includes the evolution (in moment space) of the 
parton distributions from their input scale to the scale $p_T$ 
relevant to this case. Clearly, an implementation into a 
full parton density fitting procedure is now readily possible. 

\begin{figure}[h]
\epsfig{figure=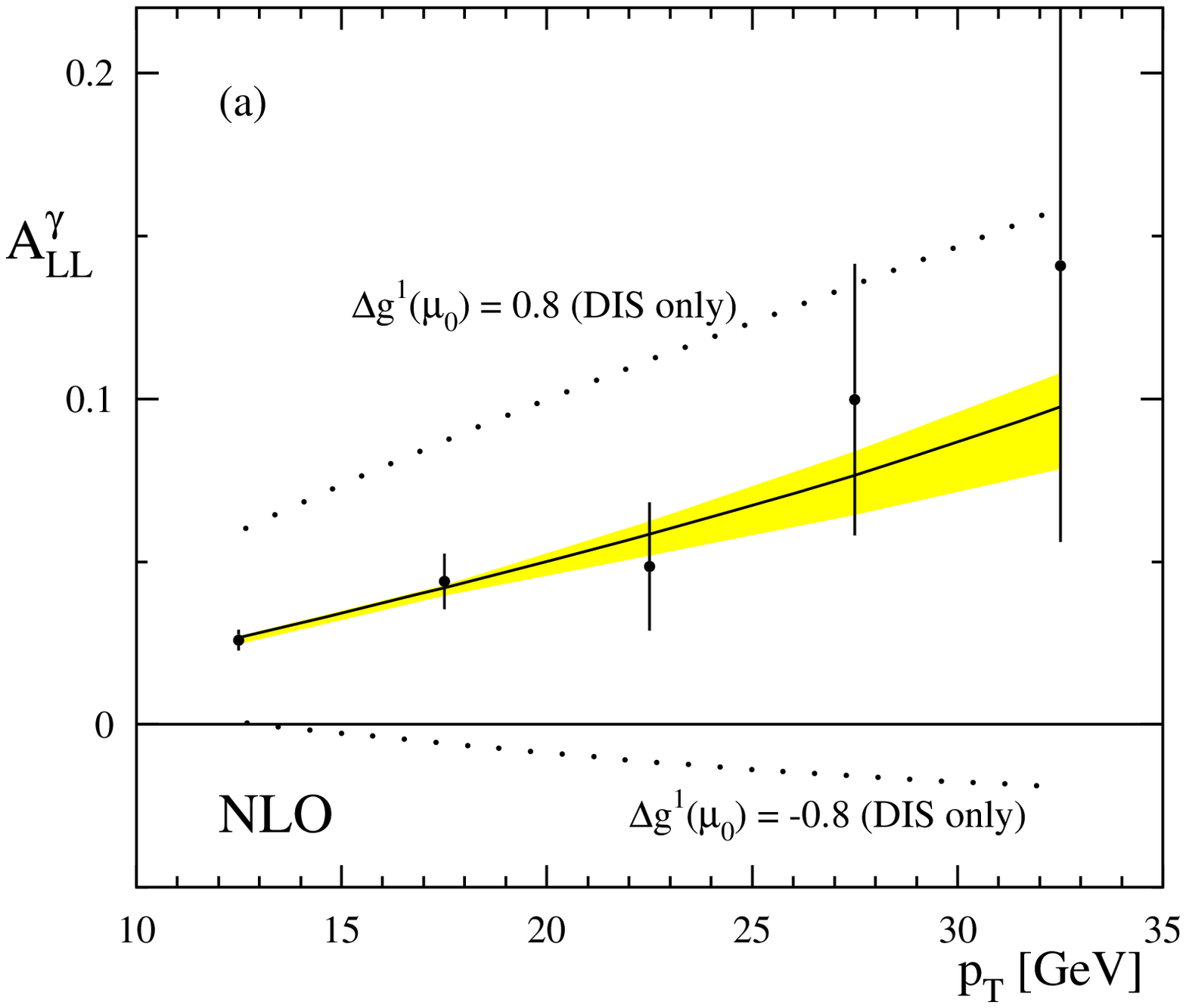,width=0.5\textwidth}

\vspace*{-0.5cm}
\epsfig{figure=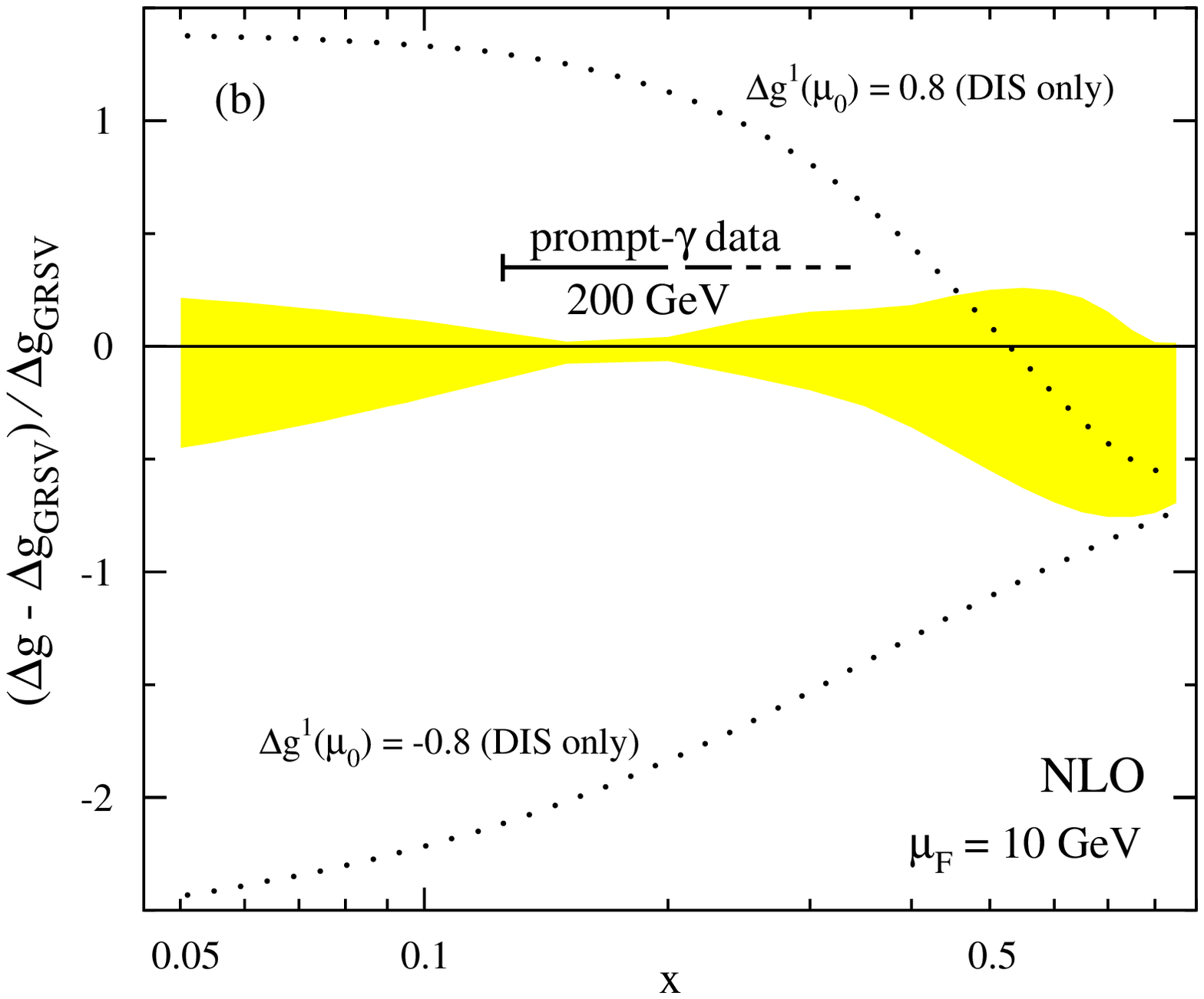,width=0.5\textwidth}
\vspace*{-1.5cm}
\caption{\sf {\bf{(a):}} Generated pseudo data for $A_{LL}^{\gamma}$
based on a calculation using the spin-dependent parton densities of \cite{grsv}
(solid line). The shaded band corresponds to the results of a large number of 
combined fits to DIS and $A_{LL}^{\gamma}$ data (see text).
{\bf{(b):}} Variations of the polarized gluon densities 
obtained in
the combined fits, with respect to $\Delta g$ of \cite{grsv}. 
\label{fig3}}
\vspace*{-0.7cm}
\end{figure}

To give an example, we finally perform a ``toy''
global analysis of the available data on polarized deep-inelastic
scattering~\cite{emc} and of {\em fictitious} data on prompt photon
production at RHIC~\cite{rhic}, which we project by simply calculating 
$A_{LL}^{\gamma}$ to NLO 
using the sets of polarized and unpolarized parton distributions 
of \cite{grsv} 
and \cite{grv}, respectively. For an estimate of the anticipated 
$1\sigma$ errors on the ``data'' for $A_{LL}^{\gamma}$, we use the 
numbers reported in~\cite{rhic}. We subsequently apply a random Gaussian 
shift of the pseudo-data, allowing them to vary within $1\sigma$. 
The ``data'', as well as the underlying
theoretical calculation of $A_{LL}^{\gamma}$ based
on the spin-dependent parton densities of \cite{grsv} (solid line), 
are shown in Fig.~\ref{fig3}(a). 

Next, we perform a large number of fits to the full,
{\em DIS plus projected prompt photon}, ``data set''. We simultaneously
fit {\em all} polarized parton densities,
choosing the distributions of \cite{grsv} as the input for the
$\Delta q$, $\Delta \bar{q}$, but using randomly chosen values for 
the parameters in the ansatz for the polarized gluon distribution at 
the input scale $\mu_0$. Regarding the details of the evolution, we stay 
within the setup of~\cite{grsv}, but we
choose a more flexible ansatz for $\Delta g$:
\begin{equation} \label{eq12}
\Delta g (x,\mu_0) = N \, x^{\alpha} \, (1-x)^{\beta} \,
(1+\gamma \, x)\; g(x,\mu_0) \; , 
\end{equation} 
which also allows for a zero.
$g(x,\mu_0)$ is the unpolarized gluon density~\cite{grv} at the
input scale of \cite{grsv}. Ideally, thanks to the strong sensitivity of the 
prompt photon reaction to $\Delta g$,  
the gluon density in each fit should return close to the function we assumed 
when calculating the fictitious prompt photon ``data'', in the region 
of $x$ probed by the data. Indeed, as shown in Fig.~\ref{fig3}(b), 
this happens. The shaded band illustrates the deviations
of the gluon densities obtained from the global fits to the 
``reference $\Delta g$'' \cite{grsv} used in generating the pseudo-data.
\vspace*{-1.2cm}
It should be stressed that only those fits are admitted
to the band that give a good simultaneous description of the DIS
{\em and} $A_{LL}^{\gamma}$ data. Here we have tolerated a maximum
increase of the total $\chi^2$ by up to four units from its minimum value.
The shaded area in Fig.~\ref{fig3}(a) shows the variations in
$A_{LL}^{\gamma}$.

As is expected, the gluon density is rather tightly constrained in
the $x$-region dominantly probed by the prompt photon data. This is true 
in particular at $x\approx 0.15$, as a result of the most precise ``data 
point'' for $A_{LL}^{\gamma}$ at $p_T=12.5\,\mathrm{GeV}$.
Fig.~\ref{fig3}(b) shows also two extreme gluon densities with first moments 
$\Delta g^1(\mu_0)=\pm 0.8$ (dotted lines), which are both in 
perfect agreement with all presently available DIS data. 
It should be noted that future measurements of $A_{LL}^{\gamma}$ at RHIC
at $\sqrt{S}=500\,\mathrm{GeV}$ and for similar $p_T$ values
would further constrain $\Delta g$ 
in the $x$-region between 0.05 and 0.1.
Although our analysis still contains a certain bias by choosing only 
the framework of \cite{grsv} for the fits as well as by our choice of
what $\chi^2$ values are still tolerable, it clearly outlines the 
potential and importance of upcoming measurements of 
$A_{LL}^{\gamma}$ at RHIC for determining $\Delta g$.

\section{Spin asymmetries and resummation}
In the unpolarized case, a pattern of disagreement between theoretical
predictions and experimental data for prompt photon production has been 
observed in recent years~\cite{abe94}, not globally curable by 
`fine-tuning' the gluon density~\cite{vogt95}. The main problems reside
in the fixed-target region, where NLO theory dramatically underpredicts 
some data sets. At collider energies, as relevant to RHIC, 
there is less reason for concern.

In view of this, various improvements of the theoretical framework have been 
developed. One of them resorts to applying `threshold' resummation 
to the prompt photon cross section~\cite{LOS}, which organizes to all orders 
in $\alpha_s$ large logarithmic corrections to partonic hard scattering
associated with emission of soft gluons. As the partonic c.m.s.\
energy $\hat s$ approaches its minimum value at $\hat s = 4p_T^2$,
corresponding to `partonic threshold' when the initial
partons have just enough energy to produce the photon and
the recoiling jet, the phase space available for gluon bremsstrahlung 
vanishes, resulting in corrections to the partonic cross section 
$d\hat\sigma/dp_T$ as large as $\alpha_s^k \ln^{2k}(1-4p_T^2/\hat s) \, \hat 
\sigma^{\rm Born}$ at $k$-th order in PT. Threshold 
resummation~\cite{dyresum,LOS} organizes this singular 
behavior of $d\hat\sigma/dp_T$ to all orders.
It is again carried out in Mellin-$n$ moment space, where the logarithms 
have the form $\alpha_s^k \ln^{2k} (n) \, \hat 
\sigma^{\rm Born} (n)$. Its application is particularly interesting 
in the fixed-target regime, since here the highest $x_T$ are attained in 
the data and the discrepancy between data and theory is 
largest.

\begin{figure}[t]
\begin{center}
\epsfig{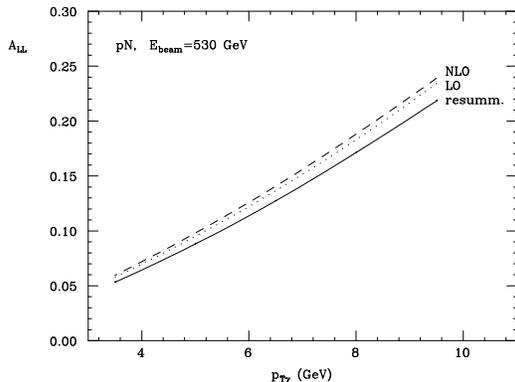}
\end{center}
\vspace*{-1.2cm}
\caption{Spin asymmetry \protect{$A_{LL}$} at LO, NLO,
and including NLL threshold resummation.
\label{fig:f2}}
\vskip -8mm
\end{figure}

In phenomenological applications~\cite{CMNOV,KO} 
of threshold resummation, 
one finds a significant, albeit not sufficient, enhancement of the 
theory prediction in the fixed-target regime at large values of 
$p_T/\sqrt{S}$, accompanied by a dramatic reduction of scale 
dependence~\cite{CMNOV,KO}. Thanks to the universal 
structure of soft-gluon emission, it is straightforward to apply threshold 
resummation to the polarized cross section. Fig.~\ref{fig:f2} shows the 
resulting effects on the {\em spin asymmetry} $A_{LL}$, for a `toy' example 
that assumes a fictitious polarized set-up of the E706 
experiment. Details are as in~\cite{CMNOV}. Even 
though Fig.~\ref{fig:f2} does not directly 
refer to the case of RHIC, it is good news that resummation effects 
cancel to a large extent in $A_{LL}$ for our present example.

A similar cancellation of higher-order effects associated with
soft-gluon emission is found for the parity-violating longitudinal 
single-spin asymmetry for jet production in $\vec{p}p$ collisions,
which has been proposed~\cite{virey,rhic} as a candidate indicator for 
physics beyond the Standard Model and is also measurable at RHIC.
 
\end{document}